\documentclass{appolb}

\usepackage{hyperref}
\usepackage{graphics}
\usepackage{amsmath}
\usepackage{amssymb,amscd}
\usepackage{graphicx}

\usepackage[all]{xy}

\def\={\ =\ }
\def\dd{{\rm d}}

\newcommand{\Gr}{\mathbb{G}{\rm r}}
\newcommand{\Fl}{\mathbb{F}{\rm l}}

\def\beq{\begin{equation}}
\def\eeq{\end{equation}}
\def\bea{\begin{eqnarray}}
\def\eea{\end{eqnarray}}

\DeclareMathOperator{\Mat}{Mat}

\DeclareMathOperator{\Ext}{Ext}

\DeclareMathOperator{\ch}{ch}

\def\ii{{\,{\rm i}\,}}

\newcommand{\bbr}{\mathbb{R}}

\newcommand{\bbc}{\mathbb{C}}
\newcommand{\bbC}{\mathbb{C}}

\newcommand{\bbP}{\mathbb{P}}

\newcommand{\cD}{\mathcal{D}}

\newcommand{\cali}{\mathcal{I}}
\newcommand{\calj}{\mathcal{J}}
\newcommand{\calo}{\mathcal{O}}
\newcommand{\calm}{\mathcal{M}}
\newcommand{\caln}{\mathcal{N}}

\newcommand{\calq}{\mathcal{Q}}

\newcommand{\sfm}{\mathsf{M}}

\def\hil{{\mathcal H}}

\def\Id{{\rm id}}

\def\ch{{\rm ch}}

\newcommand{\bbz}{{\mathbb Z}}

\usepackage{epsfig}

\input{epsf}

\begin{document}

\eqsec  
\title{CRYSTALS, INSTANTONS \\ AND QUANTUM TORIC
  GEOMETRY\thanks{Contribution to the proceedings of ``Geometry
  and Physics in Cracow'', Jagiellonian University, Cracow, Poland,
  September 21--25, 2010. To
  be published in {\sl Acta Physica Polonica Proceedings
    Supplement}.\\ Report numbers: \ HWM--11--4 \ , \ EMPG--11--05}}%

\author{Richard J. Szabo
\address{Department of Mathematics, Heriot--Watt University, Edinburgh
EH14 4AS, U.K.}}

\maketitle
\begin{abstract}
We describe the statistical mechanics of a melting crystal in three
dimensions and its relation to a diverse range of models arising in
combinatorics, algebraic geometry, integrable systems, low-dimensional
gauge theories, topological string theory and quantum gravity. Its
partition function can be computed by enumerating
the contributions from noncommutative instantons to a six-dimensional
cohomological gauge theory, which yields a dynamical realization of the
crystal as a discretization of spacetime at the
Planck scale. We describe analogous relations between a melting
crystal model in two dimensions and $\mathcal{N}=4$ supersymmetric
Yang--Mills theory in four dimensions. We elaborate on some
mathematical details of the construction of the quantum geometry which
combines methods from toric geometry, isospectral deformation theory and
noncommutative geometry in braided monoidal categories. In particular,
we relate the construction of noncommutative instantons to 
deformed ADHM data, torsion-free modules and a noncommutative twistor correspondence.
\end{abstract}

  
\section{Introduction}

A classical instanton is a connection on a smooth $SU(r)$ vector bundle $E$ over an oriented
Riemannian four-manifold $X$ with anti-self-dual curvature two-form
$F_A$, \ie
\begin{equation}
*F_A\=-F_A \ ,
\label{ASDeqs}\end{equation}
where $*$ denotes the Hodge duality operator on $X$. Such field configurations are 
labelled by their ``topological charge'', which is the instanton number
defined as the second Chern class 
$$c_2(E)\=\frac1{8\pi^2}\,\int_X\,
  \Tr(F_A\wedge F_A) \=k \ \in \ H^4(X,\bbz)$$
of the bundle $E$. (The first Chern class $c_1(E)=0$.) The
prototypical example is the case of instantons on
the four-sphere $X=S^4$. In this case, there are one-to-one
correspondences between the following classes of objects:
\begin{itemize}
\item[1.] Instantons on the Euclidean four-plane $\bbr^4$ of topological charge $k$ and finite Yang--Mills energy.
\item[2.] Rank $r$ holomorphic vector bundles $E$ over the complex
  projective plane $\bbP^2$ with $c_2(E)=k$ which are trivial on a
  projective line $\bbP^1$ at infinity.
\item[3.] Linear algebraic ADHM data.
\item[4.] Rank $r$ holomorphic vector bundles $E$ over the projective
  three-space $\bbP^3$ with $c_2(E)=k$ which are trivial on a
  $\bbP^1$ at infinity, have vanishing cohomology
  $H^1(\bbP^3,E(-2))=0$, and satisfy a certain reality condition.
\end{itemize}
The first equivalence follows since one can
glue together local connections on the northern and
southern hemispheres of $S^4$, with suitable boundary conditions at
infinity in $\bbr^4$, to produce a global instanton~\cite{Atiyah79}. The second equivalence is known as the \emph{Hitchin--Kobayashi
  correspondence} and it gives a construction of the
instanton moduli space in algebraic geometry~\cite{Donaldson84}. The third equivalence
gives an explicit construction of the instanton connections on
$\bbr^4$ through
solutions of the celebrated ADHM matrix equations~\cite{ADHM}. The fourth
equivalence yields the \emph{Atiyah--Penrose--Ward twistor
  correspondence} which can be used to explicitly construct instantons on $S^4$~\cite{AtiyahWard}.

The anti-self-duality equations (\ref{ASDeqs}) have a natural generalization to
higher-dimensional K\"ahler manifolds $X$ called the
Donaldson--Uhlenbeck--Yau equations~\cite{DUY}. Irreducible gauge
connections which solve these equations are in one-to-one
correspondence with stable holomorphic vector bundles over $X$; they
naturally arise in compactifications of heterotic string theory as the
condition for at least one unbroken supersymmetry in the low-energy
effective field theory. Of particular interest are the
cases in which $X$ is a toric manifold, like the original example
$\bbr^4\cong\bbc^2$. In this case, the torus symmetries of $X$ lift to the
instanton moduli space and the powerful techniques of equivariant
localization can be used to compute the exact instanton contributions
to the partition functions of supersymmetric gauge theories on
$X$~\cite{Nekrasov02,Nekrasov06,Nekrasov05}. Besides their intrinsic
interest as exactly solvable models which capture physical regimes of more
realistic quantum field theories, these partition functions also enumerate BPS bound states of D-branes in
Type~II string theory in certain regions of the moduli
space. Instanton counting has also found applications in geometry through the computation of enumerative invariants of
manifolds, \eg the Seiberg--Witten~\cite{Nekrasov02} and Donaldson invariants~\cite{DonaldsonKron} when
$\dim_\bbc X=2$, and the Donaldson--Thomas invariants when $\dim_\bbc X=3$~\cite{Maulik06,Iqbal08}.

The enumeration of instantons on a general toric $d$-fold $X$ in the approach
of~\cite{Nekrasov05,Iqbal08} is somewhat heuristic. It begins with the
\emph{local} enumeration of (generalized) noncommutative instantons on
each torus invariant open patch $\bbc^d \subset X$. A Moyal
deformation of these patches is simple enough to enable explicit
construction of the instanton connections in this case, whose
contributions to the partition function can then
be assembled to global quantities using the gluing rules
  of (commutative) toric geometry. This construction
  gives rise to a crystalline structure of
  spacetime, which as an integrable model of lattice statistical mechanics has many
  interesting features in its own right. We will interpret this
  crystal model as a quantization of spacetime geometry at the Planck
  scale, induced by quantum gravitational fluctuations which are
  effectively encoded in the dynamics of noncommutative
  instantons. When $d=3$ we will give a very precise
  dynamical realization of all these correspondences, while for $d=2$ we can give an explicit construction of the instanton moduli space and its associated gauge connections.

Although this heuristic picture is nice and certainly very useful, one
would like to go beyond it somewhat by finding a \emph{global} notion
of ``noncommutative toric variety'', and the construction of
instantons thereon. This would cast the picture of dynamical quantum geometry
into the more precise and rigorous framework of
noncommutative geometry. Another reason is that such varieties
naturally arise in string geometry. For example, chiral fermions on a
``quantum curve'' can be embedded in string
  theory as a collection of intersecting D-branes in a background
  supergravity $B$-field. Such a configuration is described mathematically by a
  holonomic $D$-module~\cite{Dijkgraaf09}, roughly speaking a
  representation or sheaf over an algebra of differential operators. In certain
  instances, there is an equivalence between categories of $D$-modules
  and of modules on a noncommutative variety. Our constructions give
  examples of such noncommutative varieties, and hence of the quantum
  geometries eluded to in~\cite{Dijkgraaf09}. The simplest example of
  this correspondence is between the right ideals of the Weyl algebra
  $\bbc[z,\partial_z]$, \ie the algebra of differential operators on
  the affine line, and line bundles on a noncommutative
  $\bbP^2$~\cite{Baranovsky02}. In turn, vector bundles on
  noncommutative $\bbP^2$ correspond to instantons on a noncommutative
  $\bbr^4$~\cite{NekrasovSchwarz98,Kapustin01}. Hence the construction
  of instantons on noncommutative toric varieties produces a
  sharper picture of the dynamically induced quantum geometry.

\section{Crystal melting in three dimensions}

In this section we introduce the melting crystal model and describe
its statistical mechanics. We relate it to enumerative problems in
combinatorics and toric geometry, and explain its interpretation in
string theory. We then relate the model to an integrable hierarchy and
recast it as a matrix model, which leads into our first gauge theory
characterization of the crystal in terms of Chern--Simons theory. We
defer describing the relationship with noncommutative instantons to later sections, and
begin with the three-dimensional case wherein the complete story is
best understood. 

\subsection{Statistical mechanics and random plane partitions}

\begin{figure}[h]
\begin{center}
\epsfxsize=1.5 in\epsfbox{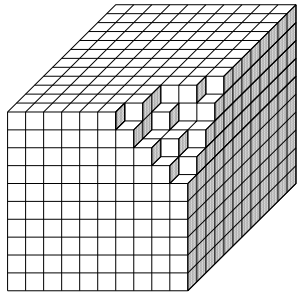}
\end{center}
\caption{Melting crystal corner in three dimensions.}
\label{crystal}\end{figure}

The model of a melting crystal corner was introduced
in~\cite{Okounkov06} and is depicted in Fig.~\ref{crystal}.
The crystal is a rectangular array of unit cubes located in the
positive octant of $\bbr^3$. It melts starting from its outermost
right-hand corner according to the \emph{melting crystal rule}: a cube
located at $(I,J,K)\in \bbz_{\geq0}^3\subset\bbr^3$ evaporates if and
only if all cubes located at $(i,j,k)$ with $i\leq I$, $j\leq J$ and $
k\leq K$ have already evaporated; this rule roughly states that an atom can
be removed only if all atoms on top of it have been removed. Removing
each atom from the corner of the crystal contributes a factor $q=\e^{-\mu/T}$
to the Boltzmann weight, where $\mu$ is the chemical potential and $T$
is the temperature.

\begin{figure}[h]
\begin{center}
\epsfxsize=1.5 in\epsfbox{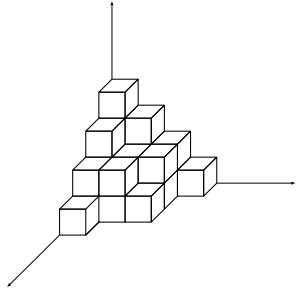}
\end{center}
\caption{A three-dimensional Young diagram.}
\label{3dYoung}\end{figure}

We can map this model onto a combinatorial problem by piling cubes in
the corner of a room as they are removed from the crystal. This is
depicted in Fig.~\ref{3dYoung}.
The melting crystal rule implies that
piling $\pi_{i,j}$ cubes vertically at
position $(i,j,0)$ gives a rectangular array of positive integers
$
\pi= (\pi_{i,j}) 
$
such that the entries of $\pi$ decrease as we move
along the rows and columns, \ie
$
\pi_{i,j}\geq\pi_{i+1,j}$ and
$\pi_{i,j}\geq\pi_{i,j+1}$.
Such an object is called a \emph{plane partition} or
\emph{three-dimensional Young diagram}.

Plane partitions generalize the notion of ordinary partition or Young
diagram; recall that this is an increasing sequence of positive
integers, $\lambda= (\lambda_1,\lambda_2,\dots)$, $\lambda_i\geq
\lambda_{i+1}\geq 0$, where $\lambda_i$ gives the length of the $i$-th
row in the associated Ferrers diagrams of the semi-standard Young tableaux $T$ of shape $\lambda$. There
are in fact two ways in
which ordinary partitions will play a role. Firstly, the ``diagonal
slices'' of a plane partition $\pi$, \eg $\lambda=(\pi_{i,i})$, define a sequence of 
ordinary partitions obeying ``interlacing relations''. Secondly, we can consider three-dimensional Young
diagrams with infinitely many boxes which freeze along each coordinate
direction to a two-dimensional Young diagram projected in the
respective coordinate plane.

The statistical mechanics of crystal melting is now defined in a
canonical ensemble in which each plane partition $\pi$ has
energy proportional to the total number of cubes $|\pi|=
\sum_{i,j\geq1}\,\pi_{i,j}$.
The canonical partition function is then the generating function for
plane partitions and is given by
\begin{eqnarray*}
Z_{\bbC^3} \ :=\ \sum_\pi\,q^{|\pi|} \= \sum_{k=0}^\infty\,pp(k)\,q^k \ ,
\end{eqnarray*}
where $pp(k)$ is the number of plane partitions $\pi$ with
$|\pi|=k$ boxes. This enumerative problem was solved long ago by MacMahon
with the result~\cite{Stanley}
\beq
Z_{\bbC^3} \= \prod_{n=1}^\infty\,\frac1{\big(1-q^n\big)^n}~=:~M(q) \
. 
\label{ZC3}\eeq
The function $M(q)$ is called the \emph{MacMahon function}. It generalizes the Euler
function which is the generating function for partitions. From the perspective of six-dimensional
gauge theory that we shall take later on, the integers $pp(k)$ count
the number of bound states of $k$ D0-branes with a single D6-brane
wrapping $\bbC^3$. Then the gauge theory with partition function (\ref{ZC3}) is dual
to topological string theory on the target space $\bbc^3$~\cite{Okounkov06,Maulik06,Iqbal08}.

This statistical mechanics model is also intimately related to the theory of symmetric functions. Given a partition $\lambda$ as above, the Schur polynomial in the variables $x=(x_1,x_2,\dots)$ is the formal power series $s_\lambda(x_1,x_2,\dots)=\sum_T\, x^T$, with $x^T:=x^{\lambda_1(T)}\, x_2^{\lambda_2(T)}\cdots$. They constitute a special basis for the algebra of symmetric functions, and are intimately connected to the representation theory of symmetric and general linear groups~\cite{Macdonald,Stanley}. Of central interest is the specialization of the Schur polynomials in $N$ variables to $(x_1,x_2,\dots,x_N)=(1,q,\dots,q^{N-1})$, which is given by the hook-content formula
\beq
s_\lambda(1,q,\dots,q^{N-1})\= q^{N(\lambda)}\ \dim_q(\lambda) \ ,
\label{hookcontent}\eeq
where $N(\lambda)=\sum_{i\geq1}\,(i-1)\,\lambda_i$ and the $q$-hook formula
$$
\dim_q(\lambda)\= \prod_{(i,j)\in\lambda}\, \frac{[N+j-i]}{[\lambda_i+\lambda_j^t-i-j+1]}
$$
is the quantum dimension of the irreducible unitary representation of $U(N)$ associated to $\lambda$~\cite{Fuchs}. Here $[n]=q^{(n-1)/2}\, \big(q^{n/2}-q^{-n/2}\big)\big/\big(q^{1/2}-q^{-1/2}\big)$ denotes the $q$-number associated to $n\in\bbz$.

The hook-content formula (\ref{hookcontent}) implies the hook-length formula which can be generalized to give~\cite{Stanley}
$$
s_\lambda(1,q,\dots,q^{N-1})\= \sum_{\pi_c}\, q^{|\pi_c|} \ ,
$$
where the sum ranges over all column-strict partitions $\pi_c$
(equivalently reverse semi-standard Young tableaux) of shape
$\lambda$, largest part at most $N-1$, and allowing $0$ as a
part. Hence the Schur specialization is a generating function for
column-strict plane partitions. For rectangular shapes $\lambda$,
there is a simple bijection between column-strict plane partitions of
shape $\lambda$ and ordinary plane partitions of shape
$\lambda$. However, there is no such simple correspondence for
arbitrary non-rectangular shapes. But the bijection does exist in the
\emph{reverse} situation for $N\to\infty$. In the limit $N\to\infty$
the hook-content formula (\ref{hookcontent}) reduces to $s_\lambda(1,q,q^2,\dots)=q^{N(\lambda)}\big/\prod_{(i,j)\in\lambda}\, [\lambda_i+\lambda_j^t-i-j+1]$, and we have
$$
\sum_{\pi_w}\, q^{|\pi_w|}\= q^{-N(\lambda)}\, s_\lambda(1,q,q^2,\dots)
$$
where the sum ranges over all weak reverse plane partitions $\pi_w$ of shape $\lambda$.

\subsection{Toric Calabi--Yau crystals\label{ToricCY}}

This model can be generalized to a large class of melting crystals in
the following way. Let $\Gamma$ be a finite trivalent planar graph, decorated
by placing a three-dimensional partition
$\pi_v$ each vertex $v$, and a
two-dimensional partition $\lambda_e$ representing the asymptotics of 
  $\pi_v$ at each edge $e$ emanating from a vertex $v$; to each external leg of the graph $\Gamma$ we assign the
  empty partition $\lambda=\emptyset$. To each vertex we assign the
  Boltzmann weight $q$ which weighs the number of boxes; each edge $e$
  also has associated to it a formal variable $Q_e$ weighing the total
  number of boxes. The partition function is obtained by summing over all possible
  decorations by partitions and reads
\beq
Z_X\= \sum_{\stackrel{\scriptstyle{\rm Young\
      tableaux}}{\scriptstyle\lambda_e}}\ \prod_{{\rm edges}\ e}\,
Q_e^{|\lambda_e|}~\prod_{\stackrel{\scriptstyle{\rm
      vertices}}{\scriptstyle v=(e_1,e_2,e_3)}}\,
M_{\lambda_{e_1},\lambda_{e_2},\lambda_{e_3}}(q) \ ,
\label{ZX}\eeq
where
\beq
M_{\lambda,\mu,\nu}(q)\= \sum_{\pi\, :\, \partial\pi=(\lambda,\mu,\nu)}\, q^{|\pi|}
\label{Mlmnq}\eeq
is the generating function for plane partitions $\pi$ with boundaries
$\lambda,\mu,\nu$ of sizes $N_\lambda,N_\mu,N_\nu$, \ie $N_\lambda$ is
the height of the plane partition (from piling cubic boxes), while
$N_\mu$ (resp.~$N_\nu$) is the extension towards the left
(resp.~right) such that beyond $N_\mu$ (resp.~$N_\nu$) the cross
section is frozen to $\mu$ (resp.~$\nu$). When these boundary integers
are non-vanishing, \ie $\pi$ is an infinite plane partition, one must
make sense of the box count $|\pi|$ through a suitable
renormalization~\cite{Okounkov06}.

This combinatorial construction has a natural geometric meaning in the
setting of \emph{toric geometry}. A complex variety $X$ is a \emph{toric
    variety of dimension $d$} if it densely contains a (complex)
  algebraic torus $T=(\bbc^\times)^d$ and the
  natural action of $T$ on itself (by group multiplication) extends to
  a $T$-action on the whole of $X$. The simplest examples are the torus $T$ itself, the
  affine space $\bbc^d$, and the complex projective space $\bbP^d$. If
  in addition $X$ is a Calabi--Yau manifold, \ie $X$ has trivial
  canonical line bundle $c_1(K_X)=0$, then $X$ is necessarily
  non-compact.

Toric varieties are of great interest because much of their geometry
and topology are described by combinatorial data encoded in a planar
\emph{toric web diagram} $\Gamma$ defined as follows. The vertices $v$
are the fixed points of the torus action on $X$, with $T$-invariant open
chart $U\cong\bbc^d$. The edges $e$ represent $T$-invariant projective lines $\bbP^1$
  joining pairs of fixed points $v_1$ and $v_2$. The variety $X$ is
  reconstructed from this data via a set of ``gluing rules'', which
  follow from the realization that the normal bundle
  determines the local geometry of $X$ near each edge, \ie near each
  $\bbP^1$, the space $X$ looks like the bundle
  $\calo_{\bbP^1}(-m_1)\oplus\cdots \oplus \calo_{\bbP^1}(-m_{d-1})$
  over $\bbP^1$ for some integers
  $m_i$ which determine the transition functions between
  neighbouring patches. The graph $\Gamma$ has external legs if and
  only if $X$ is non-compact; the external edges are then dual to
  non-compact divisors in the geometry.

This combinatorial information can be equivalently encoded in the
dual graph which defines the \emph{toric fan} $\Sigma\subset\bbz^d$ of $X$. It
consists of \emph{maximal (polyhedral) cones} $\sigma$ which are dual to the
vertices of $\Gamma$ and which define a toric open cover $U[\sigma]$ of $X$, a set
of $d-1$-cones dual to edges, and so on. One then specifies gluing
rules along adjacent faces $\sigma\cap\tau$ of cones $\sigma$ and
$\tau$. For the example of the complex projective plane $X=\bbP^2$, the fan $\Sigma$ consists of three maximal cones $\sigma_1,\sigma_2,\sigma_3$, corresponding to the three open $\bbc^2$ charts covering $\bbP^2$, with intersections between neighbouring two-cones giving the one-cones $\sigma_i\cap\sigma_{i+1}=\tau_i$ (with indices read modulo~$3$), and triple intersection the cone point $\sigma_1\cap\sigma_2\cap\sigma_3=\{0\}$. The dual graph $\Gamma$ is a triangle.

The formal power series (\ref{ZX}) enumerates the Donaldson--Thomas
invariants of the toric Calabi--Yau
threefold $X$ with web diagram $\Gamma$; from a gauge theory
perspective, this partition function counts BPS bound states of
D6--D2--D0 branes with a single D6-brane wrapping $X$ and D2-branes
wrapping the two-cycles of $X$. After the change
of variables $q=\e^{-g_s}$, the perturbative expansion of (\ref{ZX}) in $g_s$ gives
the Gromov--Witten invariants of $X$ and coincides (up to
normalization) with the partition
function for topological string theory on $X$~\cite{Maulik06}. Indeed,
the generalized MacMahon function (\ref{Mlmnq}) coincides (up to
normalization) with the
``topological vertex'' of~\cite{Aganagic05} in the melting crystal
formulation; this is proven by rewriting the sum over plane partitions
$\pi$ as a sum over ``diagonal'' two-dimensional Young diagrams $\lambda$
weighted by powers of skew Schur functions~\cite{Okounkov06}.

\begin{figure}[h]
\begin{center}
\epsfxsize=3 in\epsfbox{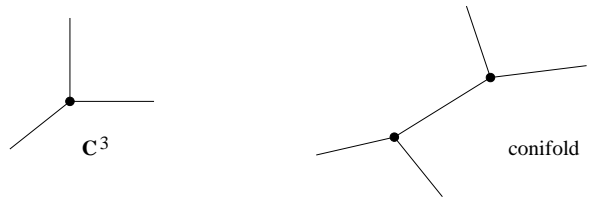}
\end{center}
\caption{Toric diagrams for $\bbc^3$ and the resolved conifold.}
\label{vertex}\end{figure}

The toric diagram for the affine space $X=\bbc^3$ is depicted on the left in
Fig.~\ref{vertex} and its partition function (\ref{ZX}) is given in
(\ref{ZC3}).
The next simplest example is the resolution of the
conifold singularity $x\,y-z\,w=0$ in $\bbc^4$, whose web diagram is
depicted on the right in Fig.~\ref{vertex}. Using the general
prescription (\ref{ZX}) one readily computes 
\begin{eqnarray}
{Z}_{\rm conifold} &=& \sum_{\lambda}\,
M_{\emptyset,\emptyset,\lambda}(q)\,
M_{\emptyset,\emptyset,\lambda}(q) \, Q^{|\lambda|} \nonumber \\[4pt] &=& \sum_{\pi_v}\,
q^{|\pi_v|+\sum_{(i,j)\in\lambda} \,(i+j+1)} \,
Q^{|\lambda|} \= M(q)^2\, M(Q,q)^{-1} \ ,
\label{Zconifold}\end{eqnarray}
where the generating function
$$
M(Q,q)\= \displaystyle{\prod_{n=1}^\infty\, \frac1{\big(1-Q\,q^n\big)^n}}
$$
counts weighted plane partitions.

In general a simple closed product formula
is not anticipated for the partition function (\ref{ZX}). They only
arise when the background $X$ has no compact divisors (or D4-branes). As the example
(\ref{Zconifold}) demonstrates, the partition function (\ref{ZX}) is
expected to contain the overall factor
$M(q)^{\chi(X)}$ enumerating degree $0$ curve classes (D0-branes)~\cite{Maulik06}, where $\chi(X)$ is the topological
Euler characteristic of $X$ which coincides with the number of
vertices in the toric diagram $\Gamma$ of $X$.

\subsection{Integrability}

The melting crystal model is an integrable system. This can be seen
through the free fermion representations of the partition functions
(\ref{ZX})~\cite{Nakatsu09,Sulkowski09}. Introduce independent
holomorphic complex
fermion fields $\psi$ and $\psi^*$ in two dimensions. In the
Neveu--Schwarz sector they have the mode expansions
$$
\psi(z)\= \sum_{m\in\bbz+1/2}\, \psi_m\,z^{-m-1/2} \qquad \mbox{and} \qquad \psi^*(z)\= \sum_{m\in\bbz+1/2}\, \psi^*_m\,z^{-m-1/2}
$$
with the non-vanishing canonical anticommutation relations
$\{\psi_m,\psi_n^*\}= \delta_{m+n,0}$.
The fermionic Fock space is built by the action of these mode
operators on the vacuum state $|0\rangle$ obeying
$\psi_n|0\rangle=0=\psi_{m}^*|0\rangle$ for $n\geq0$ and $m\geq1$. It is naturally
spanned by states labelled by Young tableaux; given a two-dimensional
partition $\lambda=(\lambda_1,\dots,\lambda_r)$ and its transpose
$\lambda^t$, one defines the basis states
$$
|\lambda\rangle\= \prod_{i=1}^r\, \psi^*_{-\lambda_i+i-1/2}\,
\psi_{-\lambda_i^t+i-1/2} |0\rangle \ .
$$

The modes $\alpha_n$ of the bosonized field $$\partial\phi(z)\= :\psi(z)\,
\psi^*(z) : \=\sum_{n\in\bbz}\, \alpha_n\, z^{-n-1} \qquad \mbox{with} \quad \alpha_n\= \sum_{m\in\bbz}\, :\psi_{n-m}\,\psi_m^*: $$ obey the
Heisenberg commutation relations
$
[\alpha_m,\alpha_n]=m\, \delta_{m+n,0}$.
They can be used to define vertex operators
$$
\Gamma_\pm(z) \= \exp\Big(\, \sum_{n>0}\, \frac{z^n}n\,
\alpha_{\pm\,n}\, \Big) \ .
$$
By using the expansion of (\ref{Mlmnq}) into Schur functions,
summed over Young diagrams, one can represent the partition functions
(\ref{ZX}) as particular vacuum correlation functions of these vertex operators. For
example, direct expansion of the infinite product in (\ref{ZC3}) gives
the fermionic representation
$$
Z_{\bbc^3}\= \langle 0|\ \Big(\,\prod_{n=-\infty}^0\, \Gamma_+(q^{-n})\,\Big)\,\Big(\,\prod_{n=0}^\infty\, \Gamma_-(q^n)\, \Big)\ |0\rangle \ .
$$
This identifies $Z_X$ as a tau-function of the
one-dimensional Toda lattice hierarchy~\cite{Nakatsu09}; the modes $\alpha_n$ play the role of ``Hamiltonians'' in the usual fermionic
formulation for tau-functions of the integrable KP and Toda hierarchies.

Natural candidates for explicit representations of
tau-functions of integrable hierarchies are provided by partition
functions of matrix models. In~\cite{Ooguri10,SzaboTierz10a} it was
shown that the expansions (\ref{ZX}) can be written as partition
functions of \emph{infinite-dimensional} unitary one-matrix models (when the underlying toric
Calabi--Yau variety $X$ has no compact divisors). For example, the
affine space partition function (\ref{ZC3}) can be expressed as the
matrix integral
\begin{eqnarray*}
Z_{\bbC^3} &=& \int_{U(\infty)}\, \dd U~ \det\Theta(U|q) \ ,
\end{eqnarray*}
while for the resolved conifold partition function (\ref{Zconifold}) one has 
\begin{eqnarray*}
Z_{\rm conifold} &=& \int_{U(\infty)}\, \dd U~\det\Big(\,
\frac{\Theta(U|q)}{\Theta(Q\,U|q)}~ \prod_{n=1}^\infty\,
\big(1+Q^{-1}\, U^{-1}\, q^n\big)\, \Big) \ ,
\end{eqnarray*}
where the elliptic theta-function is given by
$$
\Theta(u|q)\= \sum_{j=-\infty}^\infty\, q^{j^2/2}\, u^j \ .
$$
These formal expressions are defined as the $N\to\infty$ limits of the
corresponding eigenvalue integrals for the finite-dimensional unitary
group $U(N)$ with the bi-invariant Haar measure $\dd U$; the infinite unitary group here is then formally the contractible
one obeying Kuiper's theorem. In~\cite{SzaboTierz10a} these matrix
model formulas are derived straightforwardly starting from the
expansion of $M_{\lambda,\mu,\nu}(q)$ in skew Schur functions, using
Gessel's theorem to write the sum as a Toeplitz determinant, and then
using the fact that $N\times N$ Toeplitz determinants have well-known
expressions as integrals over the unitary group $U(N)$. The rank
here is infinite as we have to sum over {\it all} Young diagrams $\lambda$, with
no restrictions on the lengths of the rows $\lambda_i$, 
 in the expansion of the generating function
(\ref{Mlmnq}).

\subsection{Finite rank crystal model and Chern--Simons gauge theory}

It is natural to ask what is the meaning of the
finite rank versions of the unitary matrix integrals for the melting
crystal partition functions $Z_X$. The answer leads to
the somewhat unexpected appearence of a well-known topological gauge
theory in three dimensions. Consider Chern--Simons theory on an
oriented three-manifold $M$ with gauge group $U(N)$. The partition function is
given by the functional integral
\begin{eqnarray*}
Z_{\rm CS}^N(M) &=& \int\, {\rm D}A~\e^{\ii S_{\rm CS}[A]} \ ,
\end{eqnarray*}
where
\begin{eqnarray*}
S_{\rm CS}[A] &=& \frac k{4\pi}\, \int_M\, \Tr\big(A\wedge\dd A+\mbox{$\frac23$}\, A\wedge A\wedge A\big)
\end{eqnarray*}
for $k\in\bbz$ is the Chern--Simons action for a gauge potential $A$ of a connection one-form
on a (trivial) bundle over $M$. This gauge theory has a long history
as an exactly solvable quantum field theory which computes invariants in
three-dimensional geometric topology~\cite{Witten89}. It is given exactly by its
one-loop (semi-classical) approximation, with the partition function and Wilson loop observables localizing onto classical solutions of the Chern--Simons action, which are given by \emph{flat} connections of curvature $F_A=0$. When the three-manifold is a Seifert
fibration $M \to\Sigma$, integration over the $S^1$ fibre degrees of
freedom localizes the gauge theory onto a ``$q$-deformation'' of
two-dimensional Yang--Mills theory on the base Riemann surface
$\Sigma$~\cite{Beasley05,Caporaso06,Blau06,Griguolo07}, defined by replacing $U(N)$ representation theoretic quantities in the usual heat kernel expansion with their quantum analogs.

For example,
Chern--Simons theory on the three-sphere $M=S^3$, regarded as a circle
bundle over the two-sphere $\Sigma=S^2$ by means of the Hopf fibration
$S^3\to S^2$, is equivalent to $q$-deformed Yang--Mills theory on
$S^2$. In this case, the Chern--Simons partition function can be
reduced to an $N$-dimensional integral which is equivalent to the
Stieltjes--Wigert matrix model defined by the Hermitian matrix integral~\cite{Marino04,Aganagic04,Tierz04}
$$
Z_{\rm CS}^N(S^3)\= \int_{\mathfrak{u}(N)}\, \dd
H~\e^{-\Tr\log^2H/2g_s} \ ,
$$
where $g_s=\frac{2\pi\ii}{k+N}$. Using explicit expressions for the associated orthogonal polynomials (the Stieltjes--Wigert polynomials),
the matrix integral can be computed explicitly with the result
$$
Z_{\rm CS}^N(S^3)\= \prod_{j=1}^{N-1}\, \big(1-q^j\big)^{N-j}
$$
where
$
q= \e^{-g_s}= \e^{-2\pi \ii/(k+N)}$.
Unlike the more conventional Hermitian matrix models with polynomial potentials, this model involves an undetermined moment problem. In particular, it can equivalently be described by the unitary matrix model~\cite{Okuda05}
$$
Z_{\rm CS}^N(S^3)\= \int_{U(N)}\, \dd U~ \det\Theta(U|q) \ ,
$$
which is just the finite rank version of the unitary matrix model describing the melting crystal model on $\bbc^3$. It follows that
$Z_{\bbc^3}= \lim_{N\to \infty}\, Z_{\rm CS}^N(S^3)$,
and hence the finite $N$ crystal model may be regarded as the Chern--Simons matrix model. From the perspective of topological string theory, this correspondence is not so surprising, given that large $N$ Chern--Simons gauge theory describes the B-model dual to the A-model topological string theory on $X$. In~\cite{Ooguri10} it is shown that the spectral curve of the matrix model in the thermodynamic limit describes the mirror geometry to the A-model geometry.

This correspondence is interesting because the Chern--Simons matrix model on $M= S^3$ is known to be deeply connected to exactly solvable models of statistical mechanics and certain stochastic processes. For example, it is
related to the $N$-particle Sutherland model~\cite{SzaboTierz10b}. Moreover, in~\cite{deHaro04} it was pointed out that the matrix model expression for the Chern--Simons partition function on $S^3$ is just the extensivity property of probabilities in the Brownian motion of $N$ independent particles. A special instance of this latter connection can also be noted directly by using the observation of~\cite{Beasley05} that the Lawrence--Rozansky localization formula for $SU(2)$ Chern--Simons theory on $S^3$ amounts to rewriting the matrix model expression as
\begin{eqnarray*}
Z_{\rm CS}^2(S^3)&=& \sqrt{\frac2{k+2}}\, \sin\Big(\,\frac\pi{k+2}\,\Big)\\[4pt] &=& \frac{\e^{-\ii\pi/(k+2)}}{2\pi\ii}\, \int_{-\infty}^\infty\,\dd x\ \sinh^2\big(\mbox{$\frac12$}\, \e^{\ii\pi/4}\big)\ \e^{-\frac{k+2}{8\pi}\, x^2} \ .
\end{eqnarray*}
This is a first moment of the functional exponential of Brownian motion $A_t$ given by
$$
{\sf E}\big[(A_t)^n\big]\=\int_{\bbr}\,\dd x\ (\sinh x)^{2n}\ \frac{\e^{-x^2/2t}}{\sqrt{2\pi\,t}} \ ,
$$
where
$
A_t= \int_0^t\, \dd s\ \e^{2B_s}
$
with $B=\{B_t\ |\ t>0\}$ a one-dimensional Brownian motion~\cite{Matsumoto05}.

More intrinsically, there is a fundamental well-known connection between random plane partitions and non-intersecting lattice paths via the Lindstrom--Gessel--Viennot formalism~\cite{Gessel}. The specialization of the Schur polynomial $s_\lambda(1,q,\dots,q^{N-1})$ can be expressed as a random matrix average in the Stieltjes--Wigert ensemble~\cite{Dolivet} whose joint probability density has an interpretation as a Brownian motion. This follows from the Karlin--McGregor determinant formula for the probability measure of $N$ particles, at initial positions $\lambda=(\lambda_1,\lambda_2,\dots)$, to undergo independent Brownian motion without collision to an equispaced final position at time $t$~\cite{Karlin}. By using the Littlewood formula~\cite{Macdonald}
$$
\sum_\lambda\, s_\lambda(x_1,\dots,x_N)\=\prod_{i=1}^N\,
\frac1{1-x_i}\ \prod_{i<j}\, \frac1{1-x_i\,x_j} \ ,
$$
we can write the melting crystal partition function as a product of non-intersecting Brownian path distributions
$$
Z_{\bbc^3}\= \Big(\, \sum_\lambda\, s_\lambda(1,q,q^2,\dots)\, \Big)\, \Big(\, \sum_\lambda\, s_\lambda(-1,-q,-q^2,\dots)\, \Big) \ ,
$$
with $q=\e^{1/t}=\e^{-g_s}$.

\section{Quantization of toric geometry}

In this section we will relate the crystal melting model to the quantization of spacetime geometry. We first demonstrate how such quantization can be induced through quantum gravitational fluctuations in a certain toy model of quantum gravity. Later on we will see that this crystalline structure can be understood dynamically in terms of instantons of a topological gauge theory in six dimensions, extending the gauge theory description of the previous section, that we also describe below. We then describe the general construction of the quantum geometry, following~\cite{Cirio10}.

\subsection{K\"ahler quantum gravity}

The toy model of quantum gravity that we present was studied in the early 1990's and applies to any K\"ahler manifold in six dimensions; here we follow the presentation of~\cite{Iqbal08} (see also~\cite{Szabo10}). Let $X$ be a complex manifold of dimension $\dim_\bbC(X)=3$, with fixed nondegenerate K\"ahler $(1,1)$-form $\omega_0$ satisfying $\dd\omega_0=0$. In the following we will usually assume that $X$ is a toric Calabi--Yau threefold.

Given this data one can write down the gravitational path integral
$$
Z_X\=\sum_{[\omega]=[\omega_0]}\,
\e^{-S} \qquad \mbox{with} \quad 
S\=\frac1{g_s^2}\,\int_X\,\frac1{3!}\,\omega\wedge\omega\wedge
\omega \ .
$$
This integral is discrete; it is given by a sum over ``quantized'' K\"ahler forms $\omega$, which means that they have the same periods as the form $\omega_0$. This is tantamount to a summation over the Picard lattice $H^2(X,\bbz)$ of degree two cohomology classes of $X$, which consists of characteristic (isomorphism) classes of line bundles over $X$. Thus we decompose the ``macroscopic'' form $\omega$ into fluctuations around the ``background'' form $\omega_0$, given by the
curvature $F_A$ of a holomorphic line bundle $L\to X$, as
$
\omega=\omega_0+g_s\,F_A
$
with the fluctuation condition
$
\int_\beta\,F_A=0
$
for all two-cycles $\beta\in H_2(X,\bbz)$.

By direct substitution using the fluctuation condition, this gives the action
$$
S\=\frac1{g_s^2}\,\frac1{3!}\,\int_X\,\omega_0^3+\frac12\,\int_X\,
F_A\wedge F_A\wedge\omega_0+g_s\,\int_X\,\frac1{3!}\,F_A\wedge
F_A\wedge F_A \ .
$$
By dropping the irrelevant constant term, the statistical sum thus becomes
$$
Z_X\=\sum_{[L\to X]}\,
q^{\ch_3(L)}\ \prod_{i=1}^{b_2(X)}\,\big(Q_i\big)^{\int_{C_i}\,\ch_2(L)}
$$
where $q=\e^{-g_s}$, $Q_i=\e^{-\int_{S_i}\, \omega_0}$, $S_i\in H_2(X,\bbz)$ and $C_i\in H_4(X,\bbz)$ are dual bases of two-cycles and four-cycles, and $b_2(X)$ is the second Betti number of $X$. This partition function is of precisely the same form as the crystal partition function in the case that $X$ is toric; indeed the second Chern characteristic classes $\ch_2(L)$ of line bundles can be naturally associated to Young tableaux, while the third Chern characteristic classes $\ch_3(L)$ naturally correspond to three-dimensional Young diagrams. However, the \emph{problem} with this model as it is currently formulated is that the fluctuation condition on $F_A$ implies that all line bundles $L$ occuring in the sum are trivial,
$\ch_2(L)=\ch_3(L)=0$, and hence this model of K\"ahler quantum gravity is not well-defined.

It is the resolution to this problem that leads to the quantization of geometry.
Instead of considering smooth connections as is the usual practice, one should take $F_A$ to correspond to a {\it singular} $U(1)$ gauge field $A$ on
$X$. This procedure is well understood in algebraic geometry. It means that we should enlarge the range of the sum over line bundles to include also contributions from ideal sheaves, which fail to be holomorphic line bundles on a finite set of points, identified as the singular locus of the gauge fields. We will see later on that this extension is provided by considering the instanton solutions of gauge theory on a noncommutative
  deformation $\bbc_\theta^3$ of affine space, which are described in terms of {\it ideals}
  $\mathcal{I}$ in the polynomial algebra $\bbc[z_1,z_2,z_3]$. They correspond locally to crystalline configurations on each patch of the manifold $X$. In~\cite{Iqbal08} this phenomenon is interpreted as a gravitational \emph{quantum foam}. The gauge field configurations become non-singular on the blow-up
$
\widehat{X}\to X$
obtained by replacing the singular points with non-contractible cycles, and ideal sheaves on $X$ lift to line bundles on the resolution $\widehat{X}$; this alters the homology of $X$ and is interpreted as a spacetime topology change. In this way the molten crystal gives a discretization of the geometry of $X$ at the Planck scale; each atom
of the crystal is a fundamental unit
of the quantum geometry.

\subsection{Six-dimensional cohomological gauge theory}

A direct gauge theory realization of this construction can be given~\cite{Iqbal08,Cirafici09}, wherein one can naturally see the necessity for enlarging the space of gauge connections. The natural gauge theory on a D6-brane in Type~IIA superstring theory is a topological twist of the maximally supersymmetric Yang--Mills theory in six dimensions; the twisting carries us away from the usual physical gauge theory and is necessary to ensure supersymmetry when $X$ is curved. It can be obtained through dimensional reduction of ten-dimensional supersymmetric Yang--Mills theory over $X$, and the bosonic part of its action reads
\begin{eqnarray*}
S_{\rm bos}&=&\frac12\,\int_X\,\left(\dd_A\Phi\wedge *\dd_A\overline{\Phi}+
\big\|F_A^{2,0}\big\|^2+\big\|F_A^{1,1}\big\|^2\right) \\ && +\,
\frac12\,\int_X\,\left(F_A\wedge F_A\wedge\omega_0+
\frac{g_s}3\,F_A\wedge F_A\wedge F_A\right) \ ,
\end{eqnarray*}
where $\Phi$ is a Higgs field and $F_A=F_A^{2,0}+F_A^{1,1}+F_A^{0,2}$ is the holomorphic-antiholomorphic decomposition of the curvature two-form with respect to a chosen complex structure on $X$. The second line of this action coincides with that of the decomposed K\"ahler gravity action.

Considering the fermionic terms, the gauge theory has a large BRST symmetry, and its functional integrals (observables) localize at BRST fixed points which are given by the equations
\beq
F_A^{2,0}\=0\=F_A^{0,2} \qquad \mbox{and} \qquad
F_A^{1,1}\wedge\omega_0\wedge\omega_0 \= 0 \ .
\label{DUY}\eeq
These are the \emph{Donaldson--Uhlenbeck--Yau equations} which describe the absolute minima of the gauge theory action; the first equation says that the pertinent gauge bundle is holomorphic, while the second equation is an integrability condition on the gauge connection. Their solutions are thus BPS solutions which we interpret as (generalized) instantons. In string theory they describe BPS bound states of D6--D2--D0 branes on $X$ (in a particular chamber of the K\"ahler moduli space).

According to the general principles of cohomological gauge theory, the partition function can be computed from the localization formula onto the instanton moduli space $\sfm$ given by
$
Z_X=\int_\sfm\,e(\caln)$,
where
$e(\caln)$ denotes the Euler characteristic class of the antighost
bundle $\caln$ over $\sfm$ defined by integration over the zero modes of the antighost fields in the gauge fixed path integral. This expression is very symbolic, because the instanton moduli space is neither smooth nor even a variety. It can be made sense of using obstruction theory techniques from algebraic geometry; see~\cite{Szabo10} for a concise discussion of this point. Later on we will describe a variant of this moduli space for the four-dimensional analog of this gauge theory.

We can nevertheless formally use this Euler character formula to describe the instanton contributions to the partition function, provided we resolve at least some of the singularities of the instanton moduli space. First of all, we must deal with the non-compactness of $\sfm$. For $X= \bbc^3$, we can regularize the infrared singularities of $\sfm$ by putting the gauge theory in the supergravity ``$\Omega$-background'' introduced by Nekrasov~\cite{Nekrasov02}. This deforms the gauge theory such that the moduli space integrals can be evaluated explicitly using equivariant localization formulas with respect to the lift of the natural toric action on $X$ to $\sfm$; the torus fixed points on $\sfm$ are just the instanton gauge fields. Since $\ch_2(L)=0$ when $X$ has no non-trivial two-cycles, this saturates $Z_X$ by pointlike instantons in this case. We must also resolve the small instanton ultraviolet singularities of $\sfm$; this is achieved by replacing $X=\bbc^3\cong\bbr^6$ by its noncommutative deformation $\bbr^6_\theta$, defined by replacing the coordinates $x^i$ of $\bbr^6$ with Hermitian operators obeying Heisenberg commutation relations
\beq
\big[x^i\,,\,x^j\big]\=\ii\theta^{ij} \ ,
\label{Heisenberg}\eeq
with a constant, real-valued, non-degenerate antisymmetric deformation matrix $(\theta^{ij})$. Thus the pertinent compactification of the instanton moduli space needed to make gauge theory quantities well-defined also naturally leads to a quantization of the target space geometry; this compactification is known to correspond to adding ideal sheaves to $\sfm$.

\subsection{Cocycle twist quantization}

In the remainder of this section we spell out the details of the construction of the quantum geometry. We use the deformation procedure of~\cite{Majid95} which is tailored to deal with instances wherein there is a symmetry group acting on a class of objects that one wishes to quantize; in our case this will be the induced action of the torus group $T$ on the algebra of functions on a toric variety. However, we spell out the construction in a very general way that can be exploited in a variety of other contexts.

Let $H$ be a commutative Hopf algebra over $\bbc$ (representing the ``symmetries'' under consideration) endowed with a linear convolution-invertible unital two-cocycle $F : H\otimes H \rightarrow \bbc$. Such a cocycle is called a ``twist''. Below we use Sweedler notation for the coproduct $\Delta:H\to H\otimes H$ of $H$, $\Delta(h)=h_{(1)}\otimes h_{(2)}$, and also $(\Delta\otimes \Id_H)\circ\Delta(h)=h_{(1)}\otimes h_{(2)}\otimes h_{(3)}=(\Id_H\otimes\Delta)\circ\Delta(h)$, with implicit summations over the factors.

Given this data, we can define a new ``twisted'' Hopf algebra $H_F$ with the same coalgebra structure as $H$, but whose algebra product is modified to
\beq
h\times_F g\ :=\ F(h_{(1)},g_{(1)})~ (h_{(2)}\, g_{(2)})~
F^{-1}(h_{(3)},g_{(3)}) \ .
\label{htimesFg}\eeq
The cocycle condition ensures that this product is associative. A
complex vector space $A$ is a left $H$-comodule if it carries a compatible left coaction $\Delta_L: A \rightarrow H\otimes A$ of
$H$ on $A$; we use the Sweedler notation $\Delta_L(a) := a^{(-1)}\otimes a^{(0)}$ for $a\in A$, again with implicit summation. The category whose objects are left $H$-comodules and whose morphisms are left $H$-coequivariant homomorphisms is denoted ${}^H\calm$. Since $H$ and $H_F$ are the same as coalgebras, every left $H$-comodule is a left $H_F$-comodule and every $H$-coequivariant homomorphism is an $H_F$-coequivariant homomorphism. This implies that there is a functorial isomorphism of categories of left comodules
$
\calq_F:{}^H\calm\rightarrow{}^{H_F}\calm$,
which simultaneously deforms any $H$-covariant construction into an
$H_F$-covariant one. It is this technique of ``functorial
quantization'' that is extremely powerful and general enough to
fulfill all our needs.

As we have written it down thus far, this categorical equivalence is
trivial, because the functor $\calq_F$ acts as the identity on objects
and morphisms of the category ${}^H\calm$. However, the category
${}^H\calm$ has more structure, and the isomorphism $\calq_F$ acts
non-trivially on this extra structure, which is that of a braided monoidal
category. The monoidal structure is provided by the ordinary tensor
product of $H$-comodules, while the braiding morphism $\Psi:A\otimes B\rightarrow B\otimes A$ on ${}^H\calm$
is given by the trivial ``flip'' morphism which interchanges factors
in a tensor product, \ie $\Psi(a\otimes b) = b\otimes a$.
Writing $A_F=\calq_F(A)$ for $A\in{}^H\calm$, we can twist the flip
morphims into a new braiding $\Psi_F:A_F\otimes B_F\rightarrow
B_F\otimes A_F$ on ${}^{H_F}\calm$ given by $$
\Psi_F(a\otimes b) \= F^{-2}\big(b^{(-1)},a^{(-1)}\big)\,
\big(b^{(0)}\otimes a^{(0)}\big) \ . $$
There is also a twisting of the monoidal structure, but we do not
write it here.

Our main interest is the comodule
twisting of algebras. An algebra $A\in{}^H\calm$ is a left
$H$-comodule algebra if its product map $A\otimes A\to A$ is an $H$-coequivariant homomorphism. The quantization functor $\calq_F$ then
generates a left $H_F$-comodule algebra $A_F$ which as a vector space is
the same as $A$ but with the new product
\beq
a\cdot b \ := \ F\big(a^{(-1)}\,,\,b^{(-1)}\big)~
\big(a^{(0)}\,b^{(0)}\big) \ .
\label{newproduct}\eeq
If $A,B$ are comodule algebras in ${}^{H_F}\calm$, then so is their
  braided tensor product
  $A\,\underline{\otimes}\, B$, which is defined to be the vector
  space $A\otimes B$ endowed with the product defined on primitive
  elements by
$$
(a\otimes b)\cdot (c\otimes d) \= a\,\Psi_F(b\otimes c)\, d \ . $$
For the trivial flip braiding, this definition coincides with the
natural product induced on $A\otimes B$.

\subsection{Noncommutative toric varieties}

We now apply this functorial deformation procedure to define the
quantization of toric varieties $ X\rightarrow X_\theta
$~\cite{Cirio10}. First, we define the noncommutative algebraic
  torus $T_\theta=(\bbc_\theta^\times)^d$ using a twisting
cocycle. The algebra dual to the torus $T$ is the Laurent polynomial
algebra $H := \bbc(t_1,\dots,t_d)= A(T)$ which is generated by
monomials $t^p :=t_1^{p_1}\cdots t_n^{p_d}$ with $p \in
\bbz^d$. Since $T$ is an abelian Lie group, $H$ has the standard
structure of a commutative Hopf algebra with coproduct, counit, and
antipode given respectively on monomials by
$
\Delta(t^p)=t^p\otimes t^p$, $\epsilon(t^p)=1$ and $S(t^p)\=t^{-p}$,
with $\Delta,\epsilon$ extended as algebra morphisms and $S$ extended
as an anti-algebra morphism. The simplest
choice of twisting cocycle $F:H\otimes H\to\bbc$ is provided by the abelian twist defined on generators by
$$
F(t_i,t_j) \= \exp\big(\mbox{$\frac\ii2$}\,\theta_{ij}\big) \ =: \
q_{ij}$$
involving complex parameters $\theta_{ij}=-\theta_{ji} \in\bbc$, and extended as a Hopf bicharacter. Since
$T$ is abelian, one then easily checks that the twist factors cancel
out in the product (\ref{htimesFg}), and hence
$H= H_F$ as Hopf
algebras. 

Nevertheless, this cocycle still induces a non-trivial
twisting of the category of $H$-comodules. For example, the coproduct
$\Delta:H \rightarrow H\otimes H$ makes the Hopf algebra $H$ itself
into a comodule algebra in ${}^H\calm$, and so the cotwisted torus
has product (\ref{newproduct}) satisfying the relations
$$
t_i\cdot t_j \= F(t_i,t_j)~t_i\, t_j \= F^2(t_i,t_j)~t_j\cdot t_i \=
q_{ij}^2~t_j\cdot t_i \ .
$$
This defines the \emph{noncommutative torus
  $A(T_\theta)$} as an algebra object of the twisted category ${}^{H_F}\calm$.

We can thus quantize any quantity on which the original torus $T$
acts. Let $X$ be a toric variety with fan $\Sigma$ consisting of a set
of cones $\sigma$. We first define noncommutative affine toric varieties $\sigma\mapsto
  A\big(U_\theta[\sigma]\big)$ as finitely-generated $H_F$-comodule subalgebras of
  $A(T_\theta)$. For example, the noncommutative affine $d$-plane is
  the variety dual to the polynomial algebra $A(\bbc_\theta^d) =
  \bbc_\theta[t_1,\dots,t_d]$ with the relations $t_i\, t_j=
  q_{ij}^2~t_j\, t_i$. This noncommutative variety is called the
  ``algebraic Moyal plane''. It can be realized in fashion similar to the more
  conventional Heisenberg commutation relations (\ref{Heisenberg}) via the map
$
t_i \mapsto z_i=\log t_i$ with $[z_i,z_j]
= \ii \theta_{ij}$.
In general, the ``patches'' of the quantum toric variety $X_\theta$
are given by a quotient of the algebra $A(\bbc_\theta^d)$ by an ideal of
relations. The gluing rules of toric geometry now translate into
algebra automorphisms between affine patches
$A\big(U_\theta[\sigma]\big)$ in the category ${}^{H_F}\calm$. This
quantization thus uses the same combinatorial data as in the
commutative case, \ie the \emph{same fan} $\Sigma$, and just deforms
the coordinate algebra of each cone
$\sigma\in\Sigma$. See~\cite{Cirio10} for further details of the
explicit construction.

\section{Crystal melting in two dimensions}

In this section we describe the melting crystal model in two dimensions, following~\cite{Cirafici09a}. The natural gauge theory counterpart in this instance is the maximally supersymmetric Yang--Mills theory in four dimensions. We discuss to what extent the analogs of all correspondences for the three-dimensional crystal hold in this case; the proper understanding of these relationships would sharpen the picture of a dynamically induced quantum geometry of \emph{four-dimensional} spacetime. In this regard, the toric geometry of six-dimensional spaces (hence those which naturally arise in string theory compactifications) is singled out as special.

\subsection{Statistical mechanics and random partitions}

\begin{figure}[h]
\begin{center}
\centering
\epsfxsize=1.2 in\epsfbox{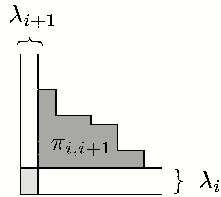}
\end{center}
\caption{Melting crystal corner in two dimensions. The index $i$ labels one-cones of the fan $\Sigma$ of a toric surface $X$, with the index pair labelling the two bounding one-cones of each torus invariant fixed point on~$X$.}
\label{crystal2D}\end{figure}

The statistical mechanics of crystal melting in two dimensions is a
combinatorial problem describing the growth of ordinary random
partitions (Young tableaux). Analogously to the three-dimensional
case, the infinite partitions label $T$-invariant open sets $U\subset
X$ of a smooth quasi-projective toric surface $X$ with asymptotics
specified by single integers along each of the two coordinate
directions. A typical configuration is depicted in Fig.~\ref{crystal2D}.
In contrast to the three-dimensional case, the vertex formalism for the counting problem greatly simplifies due to the explicit factorization of infinite Young diagrams into finite Young diagrams, as is evident from Fig.~\ref{crystal2D}. We denote this factorization symbolically as
\beq
\{\infty \ \mbox{Young tableau}\} ~ \longleftrightarrow~ \bbz_{\geq0}^2
\ \times \ \{\mbox{finite Young tableau}\} \ .
\label{Youngfact}\eeq
Geometrically this corresponds to the factorization of the Hilbert scheme of curves on $X$ into a reduced divisorial part (containing effective divisors) and a zero-dimensional punctual part (containing free embedded points); all Young diagrams (other than hook diagrams) correspond to closed subschemes of $X$ with embedded points.

The quantum version of the melting crystal corner in two dimensions is also integrable; it can be mapped exactly to the Heisenberg XXZ ferromagnetic spin chain~\cite{Dijkgraaf09a}. The classical lattice statistical mechanics on a decorated finite bivalent planar graph $\Gamma$ is described by the partition function
\beq
Z_{\rm crystal}(X) \= \sum_{\lambda_e} ~ \prod_{{\rm edges}\ e}\, G_{\lambda_e}(q,Q_e)~\prod_{\stackrel{\scriptstyle{\rm vertices}}{\scriptstyle v=(e_1,e_2)}}\, V_{\lambda_{e_1},\lambda_{e_2}}(q) \ ,
\label{crystalpart2D}\eeq
where the vertex factors are
$$
V_{\lambda_{e_1},\lambda_{e_2}}(q)\= \hat\eta(q)^{-1}\, q^{-\lambda_{e_1}\,\lambda_{e_2}}
$$
while the edge factors are given by
$$
G_{\lambda_e}(q,Q_e)\= q^{a_e\,\frac{\lambda_e\,(\lambda_e-1)}2 +\lambda_e}\, Q_e^{\lambda_e}$$
with $a_e\in\bbz$. The sum runs over $\lambda_e\in\bbz_{\geq0}$ for all internal edges $e$ of $\Gamma$, while $\lambda_e=0$ on external legs. The function $\hat\eta(q)$ is proportional to the Dedekind function $\eta(q)$; its inverse is the Euler function
$$
\displaystyle{\hat\eta(q)^{-1}\= \prod_{n=1}^\infty\,\frac1{1-q^n}\=
    \sum_{k=0}^\infty\,p(k)\,q^k}$$
where $p(k)$ is the number of partitions
    $\lambda=(\lambda_1,\lambda_2,\dots)$ of
    degree $|\lambda|=\sum_i\,\lambda_i=k$. The graph $\Gamma$ is the dual web diagram to the toric fan $\Sigma$ of a surface $X$. The integers $a_e$ are the intersection numbers between neighbouring two-cones of $\Sigma$. The appearence of the Euler function in this expression agrees with the general expectations of G\"ottsche's formula
$$
\sum_{k\geq0}\, \chi\big(X^{[k]}\big)\ q^k\= \hat\eta(q)^{-\chi(X)} \ ,
$$
where $X^{[k]}$ denotes the Hilbert scheme of $k$ points on $X$. The
six-dimensional version of this formula involving the MacMahon
function and the motivic Hilbert scheme of points was given recently
in~\cite{Behrend09}. It is
natural to ask at this stage if there exists a four-dimensional
version of ``topological string theory'' that reproduces this
counting; this point is currently under investigation.

\subsection{$\mathcal{N}=4$ supersymmetric Yang--Mills theory in four dimensions}

The relevant four-dimensional supersymmetric gauge theory is again not the physical one that appears in standard contexts such as the AdS/CFT correspondence, but rather the $\mathcal{N}=4$
  Vafa--Witten topologically twisted $U(1)$ Yang--Mills
  theory~\cite{VafaWitten94} on K\"ahler four-manifold $X$, coupled
  with instanton and monopole charges $$
k \= \frac1{8\pi^2}\,\int_X\,
  F_A\wedge F_A \qquad \mbox{and} \qquad u_i \= \frac1{2\pi}\,\int_{S_i}\,F_A
$$
for $i=1,\dots,b_2(X)$.
The topologically twisted gauge theory coincides with the physical one in the case that $X$ is a hyper-K\"ahler manifold. Under the conditions required by the Vafa--Witten vanishing theorems, the path integral localizes onto the instanton moduli space and has an expansion
\beq
\displaystyle{
Z_{\rm gauge}(X) \=
\sum_{k\geq0}\ \sum_{u\in H^2(X,\bbz)}\,\Omega(k,u)~ q^k\ \prod_{i=1}^{b_2(X)}\, Q_i^{u_i} \ ,
}
\label{VWpart}\eeq
where $\Omega(k,u)$ is the Witten index which computes the Euler
character of the moduli space of $U(1)$ instantons on $X$ (obeying the
anti-self-duality equations (\ref{ASDeqs})) with the given charges; this degeneracy factor
also counts the number of BPS bound states of D4--D2--D0 branes on $X$
(in a particular chamber of the K\"ahler moduli space).

The partition function (\ref{VWpart}) has a conjectural exact expression in the case of Hirzebruch--Jung spaces $X$~\cite{Fucito06,Griguolo07}, which are Calabi--Yau resolutions of toric orbifold singularities in four dimensions. The difficulty in making these calculations rigorous is that one needs to consider torsion-free sheaves on a ``stacky compactification'' of $X$; this variety should be a toric Deligne--Mumford stack whose coarse space is $X$~\cite{Bruzzo}. Beyond the specific examples of ALE spaces, a rigorous construction of moduli spaces of framed sheaves on these stacks is currently unknown. See~\cite{Szabo10} for further analysis of these moduli spaces.

\begin{figure}[h]
\begin{center}
\epsfxsize=4.2 in\epsfbox{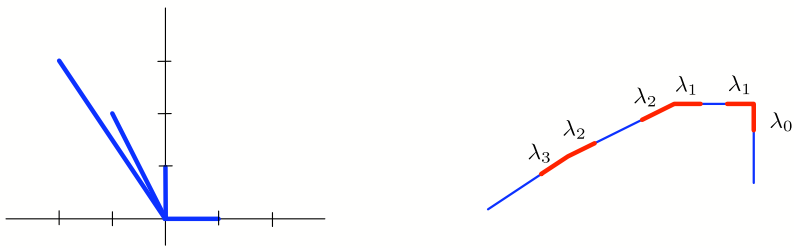}
\end{center}
\caption{Toric diagram for the $A_2$ ALE space, and its dual fan.}
\label{A2}\end{figure}

The decomposition (\ref{Youngfact}) has a gauge theory analog -- it
represents the factorization of the moduli space of rank one torsion
free sheaves on $X$ into a product of the Picard lattice of line
bundles (generated by torically invariant divisors) with the Hilbert schemes of points (ideal sheaves) on $X$. Embedding the space
of bundles with anti-self-dual gauge connections into the space of
semi-stable torsion free sheaves gives a well-defined smooth
compactification of the instanton moduli space, which is naturally
identified with a space of noncommutative instantons, as described in the next section. However, in contrast to our previous models, here the melting crystal and gauge theory problems are \emph{not} identical in four dimensions; the relation between the two enumerative problems is described in~\cite{Cirafici09a}. This can be immediately seen in the example of ALE spaces, which are resolutions of $A_n$ singularities $\bbc^2/\bbz_{n+1}$. The toric geometry for $n=2$ is depicted in Fig.~\ref{A2}.
For $n=1$ the combinatorial rules (\ref{crystalpart2D}) give the melting crystal partition function
$$ \displaystyle{Z_{\rm crystal}(A_1)\= \frac1{\hat\eta(q)^2}\, \sum_{\lambda=0}^\infty\, q^{\lambda^2}\, Q^\lambda} \ , $$ 
whereas the gauge theory instanton expansion is given by
$$\displaystyle{
Z_{\rm gauge}(A_1) \= \frac1{\hat\eta(q)^2}\, \sum_{u=-\infty}^\infty\, q^{-\frac14\,u^2}\, Q^u} \ . $$

\section{Noncommutative instantons}

In this final section we explain some details of the constructions of
noncommutative instantons and their moduli spaces, which played a
prominent role in the previous sections. We first describe the
instanton contributions to the six-dimensional cohomological gauge
theory, and demonstrate that they correctly reproduce the melting crystal
model in three dimensions. Then we analyse the instanton moduli space in four
dimensions, where an explicit construction is possible~\cite{Cirio11}.

\subsection{Noncommutative gauge theory}

Using the enveloping algebra of the Heisenberg algebra
(\ref{Heisenberg}), we regard all fields as operators on a separable
Hilbert space and turn the six-dimensional cohomological gauge theory
into a noncommutative gauge theory following the standard presciption
(see \eg~\cite{Szabo03}). For this, we represent the complex
combinations $z_a=x^{2a-1}-\ii
x^{2a}$ and $\bar z_{\bar a}=x^{2a-1}+\ii
x^{2a}$ for $a=1,2,3$ as destruction and creation operators on a three-particle Fock
space in the number basis
\beq
\hil\=\bbc\big[\bar z_{\bar1}\,,\,\bar z_{\bar 2}\,,\,
\bar z_{\bar 3}\big]|0,0,0\rangle\=
\bigoplus_{i,j,l=0}^\infty\,\bbc|i,j,l\rangle \ .
\label{Fockspace}\eeq
Introduce the covariant coordinates
$$
X^i\=x^i+\ii\theta^{ij}\,A_j \qquad \mbox{and} \qquad
Z_a\=\mbox{$\frac1{\sqrt{2}}$} \,\big(X^{2a-1}+\ii X^{2a}\big) \ .
$$
Using the Heisenberg algebra (\ref{Heisenberg}) we can represent
derivative operators as inner derivations on the noncommutative
algebra of fields; then the covariant coordinates transform
homogeneously under gauge transformations. In particular, the field strength tensor of the
gauge potential becomes a commutator of covariant coordinates, and the 
instanton equations (\ref{DUY}) become
\beq
\big[Z_a\,,\,Z_b\big]\=0 \qquad \mbox{and} \qquad \big[Z_a\,,\,\bar
Z_{\bar a}\, \big]\=3 \ .
\label{algeqs}\eeq
This is the primary technical advantage of the noncommutative
deformation -- it turns the first order partial differential equations
(\ref{DUY}) into algebraic equations.

Up to gauge equivalence, the vacuum state $F_A=0$ is given by harmonic oscillator algebra $
Z_a= z_a $. Non-vacuum solutions of (\ref{algeqs}) give fluctuations
$A_i\neq0$ around the noncommutative spacetime and hence
noncommutative instantons. The standard prescription for obtaining the
general solution starting from the vacuum field configuration is to
fix $n\geq1$, and let $U_n$ be a partial isometry 
on $\hil$ projecting out all states $|i,j,l\rangle$ with particle
number $i+j+l<n$. We
then make the ansatz $
Z_a=U_n\,z_a\,f(N)\,U_n^\dag$. The function $f(N)$ of the number
operator $N=\bar z_{\bar a}\,z_a$ is found by substituting this ansatz
into the instanton equations (\ref{algeqs}) to generate a quadratic
recursion relation for it, which has a unique solution once initial
conditions are specified; the explicit form of $f(N)$ can be found
in~\cite{Cirafici09}. The resulting instanton has topological charge
$$
k\=- \mbox{$\frac\ii6$}~ 
\Tr_\hil(F_A\wedge F_A\wedge F_A)\=\mbox{$\frac16$}\,
n\,(n+1)\,(n+2)
$$
equal to the number of states in $\hil$ with $N<n$, \ie that are removed by~$U_n$.

To identify the instanton contributions to the gauge theory partition
function, we note that $U_n$ identifies the full Fock space
(\ref{Fockspace}) with the subspace $$\hil_\cali\=
\displaystyle{\bigoplus_{f\in\cali}\,
  f\big(\bar z_{\bar 1}\,,\,\bar z_{\bar 2}\,,\,
\bar z_{\bar 3})|0,0,0\rangle} \ , $$
where $
\cali=\bbc\big\langle w_1^i\,w_2^j\,w_3^l~\big|~i+j+l\geq
n\big\rangle
$ is a monomial ideal of codimension $k$ in the polynomial
algebra $\bbc[w_1,w_2,w_3]$; it defines a plane partition
$$
\pi\=\big\{(i,j,l)~\big|~i,j,l\geq1 ~ , ~
w_1^{i-1}\,w_2^{j-1}\,w_3^{l-1}\notin\cali\big\}
$$
with $|\pi|=k$ boxes. Up to perturbative contributions from the
empty Young diagram $\pi=\emptyset$, the noncommutative instanton
contributions thus reproduce the expected MacMahon function
$Z_{\bbc^3}= M(q)$ with $q=\e^{-g_s}$. For a generic toric Calabi--Yau
threefold $X$, the corresponding field configurations are
instantons sitting on top of each other at the origin of $\bbc^3$, with
asymptotes to four-dimensional instantons along the three coordinate
axes. Patching these local contributions together then yields the three-dimensional
crystal partition function $Z_X$; see~\cite{Iqbal08,Cirafici09} for
details.

A completely analogous construction works for noncommutative
instantons in four dimensions. They sit at the origin in $\bbc^2$ and
now correspond to ordinary Young tableaux, with asymptotes along the
two coordinate axes to magnetic monopoles in two dimensions. The
associated picture of gravitational quantum foam is elucidated in
detail in~\cite{Braden04}. However,
as mentioned before, in this case the instanton contributions fail to
reproduce the two-dimensional crystal partition function.

\subsection{Instanton moduli spaces}

For the remainder of this paper we restrict to the four-dimensional case and examine the problem of constructing explicitly both the instanton moduli spaces, and the associated instanton gauge connections. For this, we compactify the affine space $\bbc^2$ to the complex projective space $\bbP^2$. The crystal partition function in this case is~\cite{Cirafici09a}
$$
Z_{\rm crystal}(\bbP^2) \= \frac1{\hat\eta(q)^3}\ \sum_{\lambda_1,\lambda_2,\lambda_3\in\bbz_{\geq0}}\, q^{\frac12\,(\lambda_1+\lambda_2+\lambda_3)^2+\frac32\,(\lambda_1+\lambda_2 +\lambda_3)}\, Q^{\lambda_1+\lambda_2+\lambda_3} \ ,
$$
while the instanton partition function is given by
$$
Z_{\rm gauge}(\bbP^2)\= \frac1{\hat\eta(q)^3}\, \sum_{u\in\bbz}\, q^{-\frac12\, u^2}\, Q^u \ .
$$

We will now construct the noncommutative projective plane
$\bbP^2_\theta$~\cite{Cirio10}. For each maximal cone $\sigma_i$ we
first construct the left $H_F$-comodule algebras
$A\big(U_\theta[\sigma_i]\big)$ dual to affine varieties which are each a copy of the noncommutative affine plane, \ie $U_\theta[\sigma_i] \cong
  \bbc_\theta^2$ for $i\=1,2,3$. The edges yield affine spaces $U_\theta[\sigma_i\cap\sigma_{i+1}]$ which are each a copy of the noncommutative projective line $\bbP_\theta^1$ dual to the polynomial algebra in two generators $w_1,w_2$ with relations
$$
w_1\,w_2\= q^2~w_2\,w_1 \qquad \mbox{and} \qquad w_1\,w_2^{-1} \= q^{-2}~w_2^{-1}\, w_1 \ ,
$$
where $q:=q_{12}$. The gluing morphisms can be summarized in the diagram
{\scriptsize{ $$
\xymatrix{
 & \bbc_\theta\big[t_1^{-1}\,,\,(t_1\,t_2^{-1})\,,\,
 (t_1\,t_2^{-1})^{-1} \big] \ar[d]&
 \\ \bbc_\theta\big[t_1^{-1}\,,\,t_1^{-1}\,t_2\big]~
\ar[ur]\ar[r]\ar[d]& ~
\bbc_\theta(t_1,t_2) ~ & ~ \ar[l]\ar[d]\ar[ul] 
\bbc_\theta\big[t_1\,t_2^{-1}\,,\,t_2^{-1}\big] \\
\bbc_\theta\big[t_1\,,\,t_1^{-1}\,,\,t_2\big] ~ \ar[ur] & ~ 
\bbc_\theta[t_1,t_2] ~ \ar[l]\ar[r] \ar[u]& ~ \bbc_\theta
\big[t_1\,,\,t_2\,,\,t_2^{-1}\big]\ar[ul]
}
$$ }}\noindent
which describes the noncommutative toric geometry of the projective plane. The Laurent algebra here is dual to the cone point of $\bbP^2_\theta$, the polynomial algebras in two variables represent the torus invariant open ``patches'', while the polynomial algebras in three generators correspond to the divisors joining patches.

To be able to proceed further, we need a more global notion of a noncommutative toric variety provided by some analog of a homogeneous coordinate algebra. In general, this is difficult to define in a manner which is compatible with the combinatorial fan construction. However, an explicit construction is possible for noncommutative projective spaces, and hence for noncommutative {\it projective} toric varieties~\cite{Cirio10}; the resulting homogeneous coordinate algebras are equivalent to those defined in~\cite{Auroux08}.

For the noncommutative projective plane, this is the polynomial algebra $A= \bbc_\theta[w_1,w_2,w_3]$ in three generators with the relations
\beq
w_1\,w_2
\=q^2~w_2\,w_1 \qquad \mbox{and} \qquad w_i\,w_3\=w_3\,w_i
\label{P2rels}\eeq
for $i=1,2$; the particular choice of $w_3$ as central element is immaterial~\cite{Cirio10}. It is graded by polynomial degree and hence $A$ defines a \emph{graded} algebra object of the category ${}^{H_F}\calm$; this grading is crucial for the ensuing constructions. Each element $w_i$ for $i=1,2,3$ generates a left denominator set in $A$, and there is a natural algebra isomorphism between the degree~$0$ left Ore localization $A[w_i^{-1}]_0$ of the algebra $A$ at each generator and the algebra $A\big(U_\theta[\sigma_i]\big)$; hence this gives an equivalent description of the noncommutative projective plane $\bbP_\theta^2$ defined above.
The graded algebra surjection $A\rightarrow A_{\infty} := A/ A\cdot w_3$ defines a noncommutative line $\bbP_\theta^1 \hookrightarrow
  \bbP_\theta^2$  ``at infinity''; it is described by setting $w_3=0$ in the relations (\ref{P2rels}).

We use the standard correspondences of noncommutative algebraic geometry. Finitely-generated graded right $A$-modules $M$ correspond to ``coherent sheaves'' on $\bbP_\theta^2$. If such a module $M$ is projective then it is thought of as a  ``bundle''. If $M$ is torsion-free, \ie it contains no finite-dimensional submodules, then $M$ embeds in a bundle. These identifications are possible and lead to well-defined constructions because the homogeneous coordinate algebras $A$ have nice ``smoothness'' properties -- they are Artin--Schelter regular algebras of global homological dimension~$3$~\cite{Cirio10,Auroux08}.

We are finally ready to construct the instanton moduli spaces
$\sfm_\theta(r,k)$, following~\cite{Cirio11}. For this, we note that
any $A$-module $M$ naturally induces an $A_{\infty}$-module
$M_{\infty}= M/M\cdot w_3$. We say that $M$ is a \emph{framed module}
if $M_{\infty}$ can be trivialized, \ie it is isomorphic to a free $A_{\infty}$-module. There is a natural notion of isomorphism for framed modules. We define $\sfm_\theta(r,k)$ to be the set of isomorphism classes of framed torsion-free $A$-modules with fixed trivialization $M_{\infty} \cong (A_{\infty})^{\oplus r}$ and with $\dim_\bbc\Ext^1(A,M(-1))= k$, where $M(-1)$ denotes the graded $A$-module $M$ with its degrees shifted by $-1$.

Torsion-free graded $A$-modules generally have natural invariants associated to them. The \emph{rank} of $M$ is the maximum number of non-zero
  direct summands of $M$; for $M\in\sfm_\theta(r,k) $ one has ${\rm rank}(M)=r$. Furthermore, the sum 
$$\displaystyle{\chi(M)\= \sum_{p\geq0}\,
(-1)^p\,\dim_\bbc\Ext^p(A,M)} 
$$
in this case is well-defined and is called the \emph{Euler characteristic} of $M$; one can show that $\chi(M)= r-k$ for $M\in\sfm_\theta(r,k) $. There is also a notion of first Chern class $c_1(M)$~\cite{Cirio11}, but we do not need this here since $c_1(M)=0$ for framed modules. However, at this purely algebraic level there is no notion of second Chern class, hence we use the Euler characteristic instead to characterize the ``instanton number'' $k$.

\subsection{Noncommutative ADHM construction}

We shall now give an equivalent characterization of the instanton moduli space $\sfm_\theta(r,k)$ in terms of linear algebraic data. Introduce matrices $B_1,B_2 \in \Mat_{k\times k}(\bbc)$,  $I \in
  \Mat_{k\times r}(\bbc)$ and $J \in\Mat_{r\times k}(\bbc)$
  satisfying the following two conditions:
\begin{itemize}
\item The \emph{noncommutative complex ADHM equation}
$$
[B_1,B_2]_\theta+I\, J \= 0 \ ,
$$
where
$$
[B_1,B_2]_\theta \ := \ B_1\, B_2-q^{-2}~ B_2\, B_1
$$
is the \emph{braided commutator} which is naturally induced by the twist quantization functor.
\item The \emph{stability} condition: There are no non-trivial invariant subspaces $0\neq V \subsetneq \bbc^k$ with $B_i(V) \subset V$ and ${\rm im}(I)
\subset V$.
\end{itemize}

Then there is a bijection between the instanton moduli space $\sfm_\theta(r,k)$ and the set of matrices $\{
  B_1,B_2,I,J\}$ obeying these two conditions modulo the natural
  action of the gauge group $GL(k,\bbc)$ given by
\beq
B_i \ \longmapsto \ g\, B_i \, g^{-1} \ , \qquad I \ \longmapsto \ g\,
I \qquad \mbox{and} \qquad J \ \longmapsto \ J\, g^{-1}
\label{ADHMaction}\eeq
for $g \in GL(k,\bbc)$. The stability condition ensures that this
group action is free and proper, hence the quotient is well-defined in
the sense of geometric invariant theory. This theorem is proven
analogously to the commutative case by constructing natural spaces of
deformed monads in the category ${}^{H_F}\calm$ on both sides of the
correspondence and proving that they are in a one-to-one
correspondence. Details can be found in~\cite{Cirio11}; there it is
also described to what extent this bijection is an isomorphism of
schemes by considering the moduli spaces as occuring in families.

\subsection{Noncommutative twistor transform}

We will close by briefly discussing to what extent the isomorphism classes in the instanton moduli space $\sfm_\theta(r,k)$ can be regarded as noncommutative gauge connections obeying some form of anti-self-duality equations. This can be achieved by constructing a noncommutative version of the twistor correspondence for instantons. We begin by defining the noncommutative Klein quadric $\Gr_\theta(2;4)
  \hookrightarrow \bbP_\Theta^5$ which is a special instance of the
  construction of noncommutative Grassmann varieties given
  in~\cite{Cirio10}. Consider the exterior algebra $\bigwedge^2\bbc^4$
  of a four-dimensional vector space which is a left $H$-comodule. It
  can be naturally regarded as a left $H$-comodule algebra, and
  accordingly the braided exterior algebra $\bigwedge_\theta^2\bbc^4$ can be regarded as an object in the category ${}^{H_F}\calm$, defined in the usual way by cocycle twist quantization. It is spanned by minors $\Lambda^J$ labelled by 2-indices $J=(j_1\ j_2)$ with $1 \leq j_1,j_2\leq4$ and satisfying the relations
$$
\Lambda^{J}\,\Lambda^{K} \= q^2_{j_{1}k_{1}}\, q^2_{j_{1}k_{2}}\,
q^2_{j_{2}k_{1}}\, q^2_{j_{2}k_{2}}~\Lambda^{K}\,\Lambda^{J} \ .
$$

There are two constraints that must be satisfied. Firstly, to ensure existence of the embedding $\Gr_\theta(2;4)
  \hookrightarrow \bbP_\Theta^5\cong \mathbb{P}(\bigwedge_{\theta}^2\bbc^4)$ we must regard these minors as homogeneous coordinates on the noncommutative projective space, which imposes the consistency condition on the deformation parameters
$$
\Theta^{JK} \= \theta^{j_{1}k_{1}}+ \theta^{j_{1}k_{2}}+
\theta^{j_{2}k_{1}}+ \theta^{j_{2}k_{2}}
$$
and hence restricts the allowed ambient varieties $\bbP_\Theta^5$
(contrary to the commutative case). Secondly, we must restrict the
homogeneous coordinates $\Lambda^J$ of the projective space to the
embedding $\Gr_\theta(2;4)$. This constraint is imposed via
noncommutative Laplace expansions of the minors, and in this case
leaves only one non-trivial relation, the quadratic \emph{noncommutative Pl\"ucker relation}~\cite{Cirio10,Cirio11}
$$
\Lambda^{(12)}\,\Lambda^{(34)} -
q_{13}\, q_{21}\, q^2_{23}\, q_{24}~\Lambda^{(13)}\,\Lambda^{(24)} +
q_{14}\, q_{21}\, q_{23}\, q_{24}^2\, q_{34}~\Lambda^{(14)}\,\Lambda^{(23)} \= 0 \ .
$$

From this noncommutative Grassmannian we can construct a noncommutative sphere $S_\theta^4$. For this, we restrict the deformation parameters to $q_{12}= q_{21}^{-1} =: q \in \bbr$ and $q_{ij}=1$ otherwise. We then define $A(S_\theta^4)$ to be the $\bbr$-algebra generated by the $*$-involution on $A(\Gr_\theta(2;4))$ given by
\begin{eqnarray*}
\Lambda^{(13)}\,^\dag\=q\,\Lambda^{(24)} \quad &\mbox{and}& \quad
\Lambda^{(14)}\,^\dag\=-q^{-1}\,\Lambda^{(23)} \ , \\[4pt]
\Lambda^{(12)}\,^\dag\=\Lambda^{(12)} \quad &\mbox{and}& \quad
\Lambda^{(34)}\,^\dag\=\Lambda^{(34)} \ .
\end{eqnarray*}

We can describe this sphere patchwise on its northern and southern hemispheres. For example, it contains a copy of the noncommutative Euclidean four-space defined as the open affine subvariety $\bbr_\theta^4 \subset S_\theta^4$
  given by the degree~$0$ right Ore localization
  $A(\Gr_{\theta}(2;4))[\Lambda^{(34)}\,^{-1}]_0$. This algebra is isomorphic to the polynomial algebra $\bbc[\xi_1,\bar\xi_1, \xi_2,\bar\xi_2]$ with generators obeying the relations
\begin{eqnarray*}
\xi_1\,\bar\xi_1\=q^2~\bar\xi_1\,\xi_1 \quad &\mbox{and}& \quad
\xi_2\,\bar\xi_2\=q^{-2}~\bar\xi_2\,\xi_2 \ ,\\[4pt]
\xi_1\,\xi_2\=q^2~\xi_2\,\xi_1 \quad &\mbox{and}& \quad
\bar\xi_1\,\bar\xi_2\=q^{-2}~\bar\xi_2\,\bar\xi_1 \ , \\[4pt]
\xi_1\,\bar\xi_2\=\bar\xi_2\,\xi_1 \quad &\mbox{and}& \quad
\xi_2\,\bar\xi_1\=\bar\xi_1\,\xi_2 \ , \\[4pt]
\xi_1^\dag\=q^{-1}~ \bar\xi_1 \quad &\mbox{and}& \quad \xi_2^\dag\=-q^{-1}~ \bar\xi_2 \ .
\end{eqnarray*}
This noncommutative four-sphere seems to be new; in particular, it is
distinct from the Connes--Landi spheres which come from isospectral
deformations~\cite{ConnesLandi01} or the quantum spheres which arise
as quantum homogeneous spaces associated to quantum
groups~\cite{Landi06}. The Connes--Landi spheres are uniquely singled
out by their cohomology, but this does not apply to complex
deformations like ours. It would be interesting to understand these spheres
in further detail, by \eg studying their cyclic cohomology and how
they are singled out as real slices inside the noncommutative Grassmannians.

The noncommutative twistor transform is constructed by means of the
twistor correspondence
\beq
\xymatrix{
 & A\big(\Fl_\theta(1,2;4)\big) & \\
A\big(\bbP_\theta^3\big) \ar[ur]^{p_1} &  & 
A\big(\Gr_\theta(2;4)\big) \ar[ul]_{p_2}
}
\label{twistorcorr}\eeq
where the algebra $A(\bbP_\theta^3)$ of the projective three-space is
the ``noncommutative twistor
algebra'', and $\Fl_\theta(1,2;4)$ is the noncommutative partial flag
variety which is most conveniently described via the braided tensor product~\cite{Cirio10}
$$
A\big(\bbP_\theta^3\big)\,\underline{\otimes} \,
A\big(\Gr_\theta(2;4)\big)~\longrightarrow~
A\big(\Fl_\theta(1,2;4)\big) \ .
$$
We will construct instantons on $S_\theta^4$ by using the
\emph{twistor transform} which is the morphism from
$A(\bbP_\theta^3)$-modules to $A(\Gr_\theta(2;4))$-modules given by
\begin{eqnarray*}
M \ \longmapsto \
  p_2{}^*\,p_{1*}(M) \qquad \mbox{with} \quad p_{1*}(M)\= \Big[
  M\otimes_{A(\bbP_\theta^3)}A\big(\Fl_{\theta}(1,2;4)\big)
  \Big]_{\rm diag} \ ,
\end{eqnarray*}
where the definition of pushforward along the correspondence diagram
(\ref{twistorcorr}) is explained in~\cite{Kapustin01}. 

The final ingredient required for the construction of noncommutative
instantons is the notion of self-conjugate instanton modules. There is
a natural quaternion structure on the homogeneous coordinate algebra
$A(\bbP_\theta^3)$ given by $\calj(w_1,w_2,w_3,w_4)
  =(w_2,-w_1,w_4,-w_3)$, which induces a functor $M \mapsto M^\dag :=
  \calj^\bullet(M)^\vee$ on the category of
  $A(\bbP_\theta^3)$-modules. On the noncommutative ADHM data it acts as $
(B_1,B_2,I,J) \mapsto(-B_2^\dag,B_1^\dag,-J^\dag,I^\dag)$. In addition
to the conditions spelled out before, let us further subject these
matrices to the \emph{noncommutative real ADHM equation}
$$
[B_1,B_1^\dag]_{\theta} +q^{-2}~[B_2,B_2^\dag]_{-\theta} +I\,I^\dag -
J^\dag\, J\= 0 \ .
$$
Then there is a bijection between the set of such matrices modulo the
restriction of the gauge symmetry (\ref{ADHMaction}) to the unitary subgroup $U(k)$, and the space of
torsion-free self-conjugate modules $M$ on $A(\bbP_\theta^3)$, \ie $M \cong
  M^\dag$, with fixed framing $M_\infty\cong(A_\infty)^{\oplus r}$ and
  $\Ext^1(A(\bbP_\theta^3),M(-2))=0$. This correspondence is again
  established using noncommutative monad techniques~\cite{Cirio11}.

The restriction of this bijection to the subvariety $\bbP_\theta^2$
gives the desired construction of anti-self-dual connections on a
canonical ``instanton bundle''. For this, we apply the twistor
transform to a self-conjugate $A(\bbP_\theta^3)$-module
  $M$, which gives a module over $A(\Gr_\theta(2;4))$. Restricting to
  the real subvariety
  $\bbr_\theta^4$ then gives the right $A(\bbr_\theta^4)$-module
$$
\caln \= \ker\cD \qquad \mbox{with} \quad \cD\= 
\begin{pmatrix}
B_1-q^{-1}\, \xi_1 &
B_2-q\,\xi_2 & I
\\ -B_2^\dag-q^{-1}\,\bar\xi_2 & B_1^\dag-q\,
\bar\xi_1 & -J^\dag
\end{pmatrix} \ ,
$$
which follows by restriction of the derived functor of the twistor
transform to $A(\bbr_\theta^4)$. By the stability condition, the map
$\cD$ is a surjective morphism of free $A(\bbr_\theta^4)$-modules such
that $\Delta = \cD\,\cD^\dag$ is an isomorphism, and the module
$\caln$ is finitely-generated and projective of rank $r$ with
  projector $P= 1-\cD^\dag\,\Delta^{-1}\, \cD$, \ie $P^2= P=
  P^\dag$.

Using the canonically defined differential structure
  $\Omega^\bullet(\bbr_\theta^4)$ obtained by deforming the classical
  calculus $\Omega^\bullet(\bbr^4)$ as a left $H$-comodule algebra
  using the twisting cocycle $F$, we obtain the \emph{instanton connection}
  $\nabla :=P\circ \dd$ in the usual sense of noncommutative
  differential geometry, 
with curvature $F_A= \nabla^2= P\,(\dd P)^2$. The difficulty at
this stage is determining what is the anti-self-duality equation that
this curvature should satisfy. In the case of isospectral
deformations~\cite{ConnesLandi01}, the Hodge duality operator is the same as in the
classical case. This is not so in our case because our deformations
are not isospectral -- the algebraic torus actions are not isometries of the
natural Riemannian structures on toric varieties. This suggests
appealing to alternative formulations of the anti-self-duality
equations (\ref{ASDeqs}). Details of all of these constructions can be found in~\cite{Cirio11}.

\section*{Acknowledgments}

The author would like to thank M.~Cirafici, L.~Cirio, A.-K.~Kashani-Poor, G.~Landi, A.~Sinkovics and M.~Tierz for enjoyable collaborations and discussions, upon which this article is based.
Various parts of this paper were presented at the workshops ``Noncommutativity and Physics:
  Spacetime Quantum Geometry'', 
Bayrischzell, Germany, May 14--17, 2010; ``Hodge Theoretic Reflections
on the String Landscape'', Edinburgh, U.K.,
June 14--18, 2010; ``Operator Algebras and Physics'', Cardiff, U.K.,
June 21--25, 2010; ``Supersymmetry in Integrable
  Systems'', Yerevan, Armenia, August 24--28, 2010; ``Geometry
  and Physics in Cracow'', Cracow, Poland, September 21--25, 2010. This work was supported by grant ST/G000514/1 ``String Theory
Scotland'' from the U.K. Science and Technology Facilities Council.

\end{document}